# A chiral 4-fermion structure for perturbed atom H: the question of antihydrogen[1].

G. Van Hooydonk, Ghent University, Faculty of Sciences, Krijgslaan 281 S30, B-9000 Ghent (Belgium)

**Abstract**
A chiral 4-fermion or 4-unit charge Coulomb system for perturbed atom H [12a] implies that an *intra-atomic* H-H-transition must obey the same quantitative criteria as a classical intra-molecular Walden inversion. In a trigonal pyramid model for a chiral molecule ABCD, with A at the top, chirality is minimal when A crosses mirror plane BCD at critical angle ½π or 90°. This critical angle is reproduced with one-electron energies of natural perturbed atom H [12a]. We prove how these results on classical 19th century classical behavior for 4-fermion atom H are still consistent with Bohr theory and the observed H-spectrum. We discuss some theoretical consequences of this intra-atomic mirror symmetry. Unlike bound state QED, we promote hydrogen mass to a critical variable for the theory and obtain, from first principles, a critical n-value, given by $n=\pi$. We detect a Mexican hat or double well potential, hidden in the observed terms of natural perturbed atom H, which reveals how the symmetry of the electron-proton bond is broken naturally. Practical consequences are that recent claims [6-9] by ATHENA- and ATRAP-collaborations on the mass-production of H may well be premature and that H-H bonding schemes may have to be reconsidered, in line with recent observations [12b].

*Introduction*

The problem of hydrogen-antihydrogen or H-H asymmetry is a longstanding and important one in theoretical physics. The question of H-H bonding is also an ultimate test for theoretical chemistry. The transition from species H to H is intimately related to the question whether or not baryon- and lepton-number is conserved [1-3]. The most intriguing aspect in a H-H-transition is probably not proton-stability or -decay but the *internal symmetry hidden* in stable H, the most abundant *neutral* species in the universe. We now try to review all internal symmetries possible in the *simple* 2 unit-charge or electron-proton bond.

If the H-H transition is a problem of internal symmetries, involving two baryons ($p^+$ and $\underline{p}^-$, eventually forming unstable antiprotonium) and two leptons ($e^-$ and $\underline{e}^+$, eventually forming equally unstable positronium), it is necessary to understand if the conversion process

$$H = p^+ + e^- \quad \leftrightarrow \quad \underline{e}^+ + \underline{p}^- = \underline{H} \qquad (1)$$

is allowed or not *in nature*. A *simple* approach consists in studying the feasibility of (1) in *empty space*, as done a long time ago by Feinberg, Goldhaber and Steigman [4]. Their treatment follows that for a muonium-antimuonium transition, put forward much earlier by Feinberg and Weinberg [5].

In particular, the *chiral* interaction $H_\chi$ for H-H is described conventionally in QFT by [4]

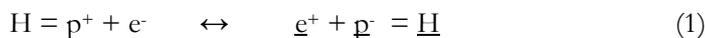

$$H_\chi = (CG_F/\sqrt{2}) \, \underline{\psi}_P \gamma_\mu (1+\gamma_5) \psi_e \underline{\psi}_P \gamma^\mu (1+\gamma_5) \psi_e \qquad (2)$$

---

[1] Based upon contributions by G. Van Hooydonk at the International Conference on Precision Physics of Small Atomic Systems (PSAS2002, Sint Petersburg, 2002) and at the Wigner Centennial (Pecs, 2002),

where C is an unknown *constant*, to be identified. The Feinberg-Weinberg approach [5] implies that, with Bohr radius r, the matrix element for $1Sp^+e^- \leftrightarrow \underline{p}^-\underline{e}^+$ due to (2) is *small*

$$\tfrac{1}{2}\delta \equiv <p^+e^-|H_\chi|\underline{p}^-\underline{e}^+> = (8CG_F/\sqrt{2})/(\pi r^3) = C\ 10^{-12}\ \text{eV}$$

$$\approx C\ 0.24\ \text{kHz} \qquad (3)$$

How small ½δ for H-H̲ (1) *really* is, depends on the value of C but also on its analytical behavior. If C<<1, the effects are of order Hz or *lower* and not yet measurable by spectroscopic methods. If C≈1, modern spectroscopy must be able to find indications for the reality of a conversion process like (1) with accurate spectral data for H. With *improbable* solution C>> 1, a problem is generated. Then, the effect of the H-H̲ conversion (1) would become comparable, if not degenerate, with what we classically describe as hyperfine and/or fine structures observed in the known H-spectrum.

*For about half a century, the problem with C in (3) is still not solved.* Using QFT result (3), spectroscopic data must, in one way or the other, be conclusive for the physics behind fundamental process (1). This brings us to *modern experimental atomic physics*, where (1) is approached differently. This *new* method envisages a large-scale production of *artificially made* H̲ ($\underline{p}^-,\underline{e}^+$), a measurement of the H̲-spectrum and, finally, a comparison with the H-spectrum. Experimentalists as well as theorists hope this method will finally reveal if there are spectroscopic differences between H and H̲ (or if CPT-symmetry holds), which will produce C in (3). This *new* method to assess (1) is reviewed in this paper as it seems to be decisive for the status of the SM and its predictions: shall we have to move towards a *new* physics or not? Following earlier attempts to produce small amounts of antihydrogen H̲ [6,7], claims were made recently that mass-production of H̲ is feasible [8,9] (see also [10]), which would open the way for measuring the H̲-spectrum. Then important term H̲1S-2S, a key datum for testing CPT or the *internal symmetries* in the H-species, can be compared with term H1S-2S, already known with great precision (parts in $10^{14}$) [11]. However, a rather embarrassing problem with *internal symmetries* within the H-species appearing in (1) is that we already detected an *intra*-atomic *mirror* symmetry effect in the line spectrum of *natural* H [12a]. This confirmed an earlier detection of the *intra-atomic* charge inversion process, present in (1), within *inter*atomic chemical bonds like $H_2$ [12b]. Analytically and unlike (2), charge-inversion (1) within bond $H_2$ corresponds simply with *an algebraic switch* (a parity operator) *within* the molecular $H_2$ Hamiltonian. This switch, not used in standard Heitler-London theory, is clearly visible in molecular band spectra [12b]. When viewed in the context of equations (1) to (3), our alternative interpretation of *available H line and band spectra* [12] suggests that C in (3) would not have to be (that) small after all.

But to verify if results [12] are really about an *overlooked* internal, *naturally occurring* (mirror-) symmetry within H, we are obliged to find additional, circumstantial, conclusive evidence, especially for result [12a], related to (1), preferentially using first principles only.

In concreto, these complementary and novel results on chirality within the electron-proton bond in H [12] must, if consistent, also be understood on the basis of classical 19th century chiral behavior, *as there is only one kind of mirror- or left-right asymmetry, whatever the constitution or complexity of the chiral system.* Such an analytical solution is not yet provided within the framework of bound state QED, as argued in [12a]. A classical solution for this longstanding problem with atomic and molecular chirality using *available* line and band spectra is therefore also decisive for claims [6-9], since the presence of H̲ in their experiments is only certain *if and only if species H̲ can be positively identified with its spectrum*. This conditio sine qua non has been the basis of important discoveries in the past. *The spectrum of artificial H̲, subject to annihilation on the spot, being unknown*, especially important term H̲1S-2S, claims [6-9] *that artificial H̲ was really produced may well have to considered as premature.*

Furthermore, priority-claim [9b] on the *first ever* assessment of the internal features of H̲ must be classified as premature too for the very same reason. But if our earlier analyses [12] were valid, details of the internal structure of H̲ have already been exposed quantitatively and with great accuracy [12a], which would overrule priority claim [9b]. Results [12] on the existence of *anti-atomic* species like H̲ in nature are not mentioned in [6-10].

From a chemical point of view, the analytical problem with H-H̲ bonding, if any bonding would occur at all, is largely dominated by the *atom-antiatom annihilation process*, following the Dirac particle-antiparticle model. Also here, a stringent reliable unambiguous result is difficult to obtain, although numerous studies exist [12c, 13], the more recent ones being inspired by the ongoing CERN-AD experiments [6-9]. The original time-line of these first chemical studies [13] almost runs parallel with those of the first physics papers [1-5], due to their common interest.

In particle physics, particle mass is probably the most important property needed in both theoretical and experimental approaches. Strangely enough, in Bohr theory, as well as in bound state QED, hydrogen mass $m_H$ does not show in the equation for its energy levels, except for a small recoil correction (of order 1/1836). In this work, we restore the first principle's importance of hydrogen mass $m_H$ for understanding the internal symmetry within H, which we can assess with a very simple model. For atom H, the electron-proton bond, we define $m_H$ ($=m_e+M_p$) explicitly as its *self-energy* (c=1). We further describe the neutral H-structure as a *static* 1D line segment r with 2 unit-charges $e^+$ and $e^-$ at its vertices. We compare our analytical results with those based upon a *dynamic* 2D (circular) model like Bohr's. In essence, we built upon the difference between the angular (2D) and radial (1D) velocity of a fermion and, in particular, on the neglect of $m_H$ in earlier bound state theories to arrive at equilibrium conditions within the H-structure. We remind that for scaling an angular velocity (circular orbits), the standard scale factor is a (simple) function of π, whereas for a radial velocity, due to a field, this is not obvious at all. For instance, we are used to scale the Bohr H-model completely with 2π (as in the de Broglie equation). *But if the field must not be scaled by π*, it can reasonably be expected that deviations may occur from the circular model. Moreover, these deviations will probably get larger, *when a critical parameter in the circular model deviates from the value of π*. Sommerfeld, who extended the Bohr model towards elliptical orbits by introducing the valuable secondary quantum number, solved part of this problem. Nevertheless, he was unable to find *a formal connection between the deviations from Bohr theory with irrational number π*. By extension, a similar remark applies to bound state QED, which is a further extension of Sommerfeld's model and can conveniently be called a Bohr-Einstein-Sommerfeld-Schrödinger-Dirac model.

(i) *Left-right asymmetry in nature*

The issue at stake with recent experiments [6-9] is whether or not it is necessary to produce an *artificial* mirror system of conventional H ($e^-$, $p^+$), i.e. anti-H or H ($e^+$, $p^-$), to get at the mechanism behind transition (1). *Left-right asymmetry is the typical end result of a symmetry breaking process as it occurs in stable, natural many particle structures*. Only *the absence, not the presence,* of symmetry determines the chirality status of a system. *Parity is not always conserved in nature*: stable 3D structures may be classified as right, positive, up, even… whereas very similar but slightly different stable 3D structures have to be classified as left, negative, down, odd…. If both forms are observed separately (or are both stable), left never matches right exactly and *parity P is not conserved*. If, at the same time, C is inverted (like in the H-H̲ transition according to a Dirac scheme), symmetry CP would not be conserved.

The advantages of a generic approach are obvious. In terms of mathematics or (quantum) number theory, there is no symmetry either between the members of a pair of consecutive integers, an odd (even) and an even (odd) integer. If number 1 is the unit (the indivisible ατομος), even integer N can be bisected *exactly*, odd integers N-1<N and N+1>N *cannot*. Only their arithmetic average N±½ is also *achiral* (less *chiral*), since it has lost the standard odd-even difference. A physics-independent and generic generalization of left-right or odd-even asymmetry is discussed in detail in [14a]. The left and right vertices of a *linear* 1D line segment [14a] like in our 1D model for neutral stable atom H must therefore already have built in generic chiral properties [14a], *experimentally visible or not*. Then *the symmetry of achiral H (or average N±½) can be broken* [14b], which secures that species H can also take the morphology of a left- or a right-handed (odd or even) structure pending certain conditions. *In this work, we quantitatively analyze in particular the environmental conditions linear H is confronted with when it is observed, i.e. when its spectrum is being measured* (a generic symmetry breaking scheme is given in [14b]).

*In the hypothesis that neutral natural 1D H can behave in chiral way indeed*, both its left- and right-handed forms (H and H̲) may be observed *in nature* too. Under these conditions, (1) would be a natural and allowed transition process but should be rewritten as

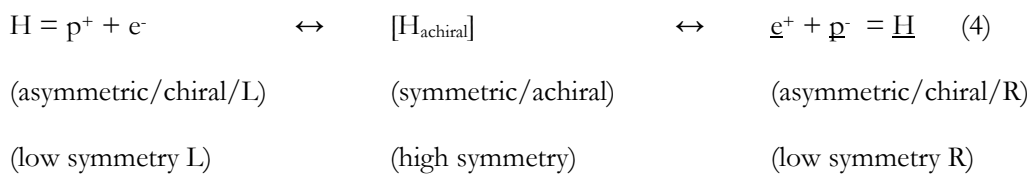

$$H = p^+ + e^- \quad \leftrightarrow \quad [H_{achiral}] \quad \leftrightarrow \quad \underline{e}^+ + \underline{p}^- = \underline{H} \quad (4)$$

(asymmetric/chiral/L)      (symmetric/achiral)      (asymmetric/chiral/R)

(low symmetry L)           (high symmetry)          (low symmetry R)

whereby an *intermediary achiral* (*more symmetrical*) state or structure for H appears. Whereas (1) refers to the discrete symmetry between H and H̲, (4) refers to a *continuous* more classical chirality model, whereby even C in (3) should not even be a *constant* (see also below). *Breaking an apparently continuous symmetry is of uttermost importance in particle physics* (e.g. the Higgs mechanism). The problem is then how to detect the existence of *chiral H-enantiomers* (left or right, H or H̲) and to find *evidence* for the presence of, and to identify the corresponding intermediary *achiral* state H$_{achiral}$, if it exists.

If *classical* mirror symmetry is involved in (4), i.e. when H̲ is really the mirror image of H, a mirror plane, *representing the achiral state of H*, must be crossed to go from H to H̲. The *only generic and exactly known* *property of a mirror plane between two mirrored objects is that it is positioned at an angle of ½π or 90°* [12a]. This simple picture provides at least with one basic exactly known parameter for system (4), i.e. angle ½π or 90°, which does not show explicitly for conversion (1). The model (4) above is generic. In a 3D Cartesian reference frame x, y, z the 2D mirror plane is x, y. The z-axis contains *the chirality determining semi-axes ±z*, corresponding with the left- or right-handed character of the 3D Cartesian reference frame. This justifies the use of a linear 1D model (on the *chiral* or field z-axis) for H as in this work and, in particular, scheme (4).

As shown in the Introduction, QFT can cope *theoretically* with chirality through (2) and (3) for process (1), but remains unsuccessful for constant C in (3). Parity-violation in particles, detected half a century ago, may be *theoretically* understood within the framework of QFT, the search for *parity-violating energy differences* $E_{PV}$ is going on strongly for a variety of systems (for atoms, see [15] and for molecules, see [16-19]). For particle physics, *symmetry breaking* in the context of the SM (and beyond) was set out by Higgs and others.

Although in chemistry, classical 19th century chiral symmetry is still the basis of many modern theoretical and practical achievements, this simple, trustworthy and very reliable procedure seems to have been more or less neglected in physics, for trying to understand the possible chiral behavior of (elementary) particles. Despite considerable efforts, a difficulty that remains is, how to *recognize classical 19th century chiral behavior for simple or light atomic systems* in the framework of QFT (2), as remarked in [12a]. This difficulty is illustrated by the fact that extremely complicated experiments like [6-9] must, eventually, come to the rescue to reach a conclusion on symmetry breaking effects in the lightest atom of all H. The basis of [6-9] is that the *man-made* mirror image of normal H (e⁻,p⁺ or electron-proton) must be H (e⁺,p⁻ or positron-antiproton) but there is uncertainty about the real existence or the stability of form H as a natural system. For instance, it is expected that $m_H = m_{\underline{H}}$. This view on (1) goes back to Dirac's particle-antiparticle model. Just like process (1), it denies the possibility that nature itself may be more imaginative and provide with a (*yet invisible*) mechanism like (4) to arrive at an intra-atomic mirror symmetry effect for the stable *2 unit-charge bond in H*. The problem is *not* how to account for this intra-atomic switch from first principles but how to put this switch quantitatively in line with classical chiral behavior as in (4) and how to put all this to the test *experimentally*, for instance by identifying the corresponding Mexican hat curve. Bound state QED as well as QFT cannot yet achieve this for H̲ [12a]. In terms of simple process (4) instead of (1), the problem for a natural species H is that it may exist in two chiral forms (*H-enantiomers*), one left- or $H_L$ the other right-handed or $H_R$ with an *achiral* state in between.

*This is why, in this work, we discuss the challenging hypothesis of H̲ exhibiting classical 19th century chiral behavior, as announced in* [12a]. We have a new look at chiral and achiral H̲ in (4) using classical criteria for chiral behavior, set out by pioneers like Pasteur, Le Bel, Van 't Hoff, Lord Kelvin, Walden…many years ago, *even before the advent of any quantum theory*. We cannot exclude the possibility that, at the time and in the aftermath of Bohr's old $1/n^2$ atomic quantum theory, something elementary and simple about classical chirality was overlooked, misinterpreted or not properly recognized as such [12a]. *An example may be the neglect of hydrogen mass $m_H$ in the theory as remarked above.*

In an older seminal work, of major importance for (4) and for the study of left- and right-handed stable structures (enantiomers), also in the context of a Higgs-mechanism, Hund [20a] showed that the potential energy curve (PEC) to describe the complete structure must be *of Mexican hat or double well-type*. The minima for the left- and right-handed structure are separated by a maximum in between. The detection of such a curve (usually a quartic) for continuous process (4) can be considered as conclusive evidence for the occurrence and the reality of left-right or mirror symmetry in an observable system/structure. *Such curves very elegantly illustrate most of the general features of symmetry breaking processes and assessing them analytically/quantitatively is an important issue in theoretical physics.*

(ii) *Constraints and problems for generating classical 19th century left-right asymmetry in simple systems like H̲. Accuracy issue.*

To deal with 19th century *classical chiral behavior*, we need at least 4 particles in a stable 3D structure, *completely devoid of any internal symmetry*. It is well known that a really chiral 3D molecule ABCD with a trigonal pyramid structure (with A at the top) shows a left-right transition, which can be described with a classical *Walden inversion*, resembling (4) [20b] or

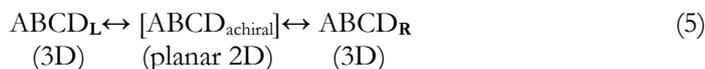

       $ABCD_L$ ↔ [$ABCD_{achiral}$] ↔ $ABCD_R$                (5)
       (3D)     (planar 2D)    (3D)

Form **L** can only go over in form **R** when top atom A passes through a mirror plane, situated at some critical point in the system, say at a value $n_0$ of a system variable n, giving an extremum at [20b]

$$n_0 = \tfrac{1}{2}\pi \text{ or } 90° \tag{6}$$

*corresponding indeed with the generic angle of a 2D mirror plane* in any 3D Cartesian reference frame as argued in (i) and in [12a,20b].

The reality of critical $n_0$ (6) was proven by Avnir's group [20b], who indeed describes chiral molecule ABCD as a trigonal pyramid with top-atom A either at the left or at the right of a triangular basis, the *perpendicular* mirror plane (angle 90°) formed by BCD. This is simply a creation of 3D Cartesian reference frames for ABCD. In this *classical* approach, atom A moves on the chiral z-axis from position $-z_L$ to $+z_R$, with z the 1D parity axis.

Apart from (6), the next quantitative criterion to be obeyed by *classical chiral behavior* is that a Mexican hat type PEC must exist, in line with Hund's early analysis [20a] (and/or with the Higgs mechanism). Classical chiral chemistry proves that also stable chiral structures containing H, say HBCD, rely on the same *phenomenological* distinction between two H-enantiomers $H_L$ and $H_R$ (H and H̲) as in (4). Since atoms BCD in HBCD form a *pseudo-mirror plane*, there must be a physical or mathematical difference between $H_L$ and $H_R$ (or H and H̲), just like that between the +z and –z semi-axes in (i). This is the reason why we use a linear 1D structure for H, although this generates some problems (see [14b] for a parallel discussion).

(a) Firstly, applying Walden inversion (5) for a 2-unit charge system in *unperturbed* H with just 2 fermions or 2 unit-charges is exactly 2 fermions (2 unit-charges or a boson) *short to arrive at a 4-particle system*.

(b) Secondly, the *artificial* H̲ experiments [6-9] may have overlooked the possibility that this Walden-type classical chiral behavior (4) or (5) might apply for a 4-fermion version of H [12a], *if it existed*. If so, there must be evidence in the *already observed and known* H-spectrum, pointing to the existence of the two enantiomers H and H̲, which was exactly our point [12a]. *Then, result (6) must interfere to get at a mirror symmetry effect in 3D*.

(c) The third problem is related to the second. If species H, when perturbed, can show chiral behavior, a Mexican hat- or double well type- potential must exist. *To my knowledge, this has never been reported before for the H species, not even by QED/QFT (but see* [20c-d] *and further below).*

(c) A fourth problem is related to the uncertainty about constant C in (3), which can be translated, for instance, in an accuracy requirement for spectral data. We expect these must be fairly *accurate* to detect mirror symmetry effects but, in reality and due to C in (3), we do not know exactly how accurate the data have to be. This uncertainty explains why in [12a] we *cautiously* proceeded with the *accurate* but *theoretical* Erickson QED-data [21]. We show below that *even less accurate experimental data* like Kelly's [22] can expose the very same intra-atomic mirror symmetry in H as that detected in [12a].

This fourth problem about the effect of the magnitude of C on accuracy constraints can be translated in terms of the physics of the system. In fact, the magnitude of C reflects the value of the asymptote for the system, to which chiral effects are linked. Bohr-Sommerfeld as well as QED theory use a shift of the primary asymptote $R_\infty$ ($\approx 109737$ cm$^{-1}$) by $(m_e/M_p)R_\infty$ ($\approx 59$ cm$^{-1}$) to account for classical recoil. Sommerfeld-Dirac-QED theories use another *extra* shift of order $\alpha^2 R_\infty$ ($\approx 5.8$ cm$^{-1}$) to account for *relativistic* effects in the first place. Many other (smaller) asymptote corrections of general combined type $(m_e/M_p)^x \alpha^y R_\infty$ contribute to the high accuracy claimed by QED-calculations [21, 23].

(d) A fifth problem is the impact of Bohr theory on the scaling of the constants or on metrology at large [24]. We will show below how important the scaling issue really is as soon as a transition from a 2D (circular orbits) to a 1D model (line segments) is at stake. A notorious feature of Bohr theory and bound state QED alike is that the most important physical characteristic of the hydrogen atom, its mass $m_H$ (= 1837.1526675 $m_e$) is absent from the energy formulae, except for a small recoil correction. *As remarked above, one would normally expect that hydrogen mass would play a very predominant role in any theory for atom H.*

(e) The last and more fundamental problem is that, in general, 4-particle systems are insoluble. Hence, *phenomenological* procedures, like in [12] must be followed.

(iii) *A 4-unit charge structure for perturbed H: energy considerations and critical n-value for circular orbits (ns-states).*
    *Introducing hydrogen self-energy $m_H c^2$ in the theory for atom H.*

Theoretically, there are a number of mathematical possibilities to arrive at a still neutral structure with 4 unit-charges, even for neutral *unperturbed* H (see for instance [14b]). For each theoretical possibility, a number of hypotheses must be made, which are beyond the scope of this work. Here, we take advantage of the fact that, in one way or the other, a *perturbed H species is created as soon as one tries to measure its spectrum.* In this hypothesis, the 2 fermions (unit-charges) of linear H with absolute total mass $m_H c^2$ (the H self-energy) are perturbed by radiative field $h\nu$, which allows us, eventually, to measure the *resonant* lines, as soon as the two interacting systems are commensurate. As suggested in [12a], our point is that, *while one is measuring the H-spectrum,* an intermediary 4-fermion complex C or structure $F_{4H}$ is formed. Then the *chiral* characteristics or the (a)symmetry of this intermediate 4-fermion complex $F_{4H}$ must be assessed. The disadvantage with 4-particle systems being that they are insoluble, there is no *ab initio* solution to calculate the energy[2].

At this stage, we explicitly introduce the self-energy of the hydrogen atom, equal to $m_H c^2$, which allows us to *approximate* the total energy $E_C$ of this complex in *perturbed* H with the sum of the energies of its components

$$E_C = E_{H(perturbed)} = E_{H(unperturbed)} + h\nu = -m_H c^2 + h\nu \qquad (7a)$$

with $E_{H(unperturbed)} = -m_H c^2$. The advantage of (7a) over Bohr theory and bound state QED is *that hydrogen mass is used as a starting point in the analysis* which seems only normal, if not even trivial, in any classical attempt to try to describe atom H as an elementary particle system. Nevertheless, we cannot solve (7a) but we can compare the magnitude of its two terms[3]. First result $\nu = m_H c^2/h$ (or a Compton length for H), is discussed elsewhere [14]. Here, we apply standard transformations

$$h\nu = hc/\lambda = (2\pi/\alpha)e_1 e_2/\lambda \qquad (7b)$$

$$m_H c^2 = s e_3 e_4/r_0 \qquad (7c)$$

$$s = 1 \text{ or } s = \tfrac{1}{2} \qquad (7d)$$

The only uncertainty concerns scale factor s (7d) but it can only take the values 1 and ½. The radiative field (7b) is described with two *mass-less* fermions, one unit charge at each of the 2 vertices of line segment $\lambda$. *Unperturbed* H (7c) has 2 unit charges at two vertices of line segment $r_0$. Competing 1D line segments $\lambda$ and $r_0$, are sufficient to describe perturbed natural neutral H. All we can say is that the structure generated for intermediate complex $F_{4H}$ can theoretically vary from linear 1D, to planar 2D and, eventually, to 3D. But if *resonance* occurs between the two separate interacting 2-unit charge Coulomb systems in (7a), we expect a standing wave equation *when the two competitors in (7a) are of equal magnitude*

$$(2\pi/\alpha)e_1e_2/\lambda = m_H c^2 = se_3e_4/r_0$$

$$(2\pi/\alpha)(r_0/\lambda)(1/s)(e_1e_2/e_3e_4) \approx 1 \qquad (7e)$$

---

[2] We cannot enumerate here the many attempts to get at a solution: the best-known attempt can be found in the theory of the chemical bond.

[3] An analytical treatment of the concurrent scale factors for the two terms in (7a) is in [25] but is irrelevant here.

Now everything depends on how H properties are scaled (an important metrological problem [24]). A critical scale factor is mass $m_H$: do we scale the perturbed H atom further with $m_H$ or do we convert $m_H$ in $m_e$, the electron mass? Fortunately, the two are simple correlated by

$$r_0 = (m_e/m_H)r_e \qquad (7f)$$

For $s=½$, substituting (7f) in (7e) and using (7d) gives

$$(4\pi/\alpha)(m_e/m_H)(r_e/\lambda)(e_1e_2/e_3e_4) \approx 1 \qquad (7g)$$

whereas for $s=1$

$$(2\pi/\alpha)(m_e/m_H)(r_e/\lambda)(e_1e_2/e_3e_4) \approx 1 \qquad (7h)$$

which are sufficiently precise and *quantitative* conditions to arrive at resonance.

With $1/\alpha = 137.03599976$ and $M_p=1836.1526675$ [26], or $m_H=1837.1526675$, we easily verify that

$$(4\pi/\alpha)(m_e/m_H) = 0.937345 \approx 1 \qquad (7i)$$

For the 4 unit-charge ratio in (2g) for $F_{4H}$ we can reasonably expect that

$$e_3e_4/e_1e_2 \approx 1 \qquad (7j)$$

With (7i) and (7j), a solution is forced upon the standing wave equation we need for resonance in intermediary complex $F_{4H}$, since substituting (7i) and (7j) directly gives a *de Broglie* type standing wave equation

$$\lambda = r_e/s \qquad (7k)$$

giving either $\lambda = r_e$ (radius) or $\lambda = 2r_e$ (diameter) pending the value of scale factor s, 1 or ½ in (7d).

*Unlike all previous theories (Bohr, Sommerfeld, Dirac, QED…), this result is obtained just by promoting the self-energy of atom H to the status it really deserves: a key element needed in the description of atom H.*

(iv) *Comparison with the de Broglie relation. The generic n(π) relation for 4-fermion H̲*

Our analysis uses linear 1D line segments, whereas Bohr's model is scaled in function of circular 2D orbits (circumferences instead of line segments, either radius or diameter, pending the s-value). As mentioned in the Introduction, scale factor π for circular systems must not be necessarily a scale factor for the field [14a]. To visualize this procedure, let us use the de Broglie standing wave equation at the roots of wave mechanics but essentially a recipe to describe resonance for circular systems

$$2\pi r_e = n\lambda \qquad (7l)$$

This relation uses 2D $2\pi r_e$ (*circumference* of a circular orbit, related to an *angular velocity*), instead of a 1D line segment (*radius or diameter* of a circular orbit, related to a *radial velocity*, a field effect [14a]). The evident consequence is that comparing our 1D result (7k) with the de Broglie 2D equation (7l) forces a very simple solution *for quantum number n in function of π or a n(π) relation*, not visible in Bohr theory (and not in QED or in QFT either). In fact, substituting (7k) in (7l) produces a critical value for n, since

$$n/s = 2\pi \qquad (7m)$$

If we now adhere to s=½, compatible with equation (7i), result (7m) finally reduces, *for circular orbits in line with the de Broglie equation*, to

$$n = \pi \qquad (7n)$$

as a critical n-value to obtain resonance in the $F_{4H}$ complex for perturbed H̲. *Result (7n) is significant in its own right*, for the reasons given in the Introduction. Unlike Bohr's equations (or those of bound state QED) or the de Broglie equation, our hypothesis of a 4-fermion structure for perturbed H̲ produces, *for the circular orbits of Bohr theory only*, a *hidden* critical value $n_0$-value (7n), invisible in integer n-based Bohr $1/n^2$ theory or in QFT. The reason is that, for this type of circular orbits, scale factor $2\pi$, is not used to scale n but to scale the quantum of action h for the angular velocity. This gives the standard *reduced* Planck constant

$$\hbar = h/2\pi \qquad (7o)$$

instead of unreduced h.

*The conclusion must be that, for the circular orbits of Bohr-Sommerfeld-de Broglie theory, i.e. ns-states, there is a direct linear connection between n and π as in (7n). For non-circular orbits, the elliptical ones in Sommerfeld's theory like np-states, this direct linear connection (7n) will probably be lost.* The simplest way to verify if this result (7n) is

consistent with experiment is to analyze one-electron energies for H ns-states, i.e. the Lyman series [12a]. One-electron energies for np-states should not obey (7n). As a simple first consequence, *this may lead to a classical and generic explanation for the standard Lamb shift as argued in* [12a].

These general *static* contours for 4-fermion complex $F_{4H}$ in perturbed H are model-independent, since only (7a) is used as a starting point. The physics behind these equations is dealt with elsewhere [14,25]. As remarked in [12a], (7i) shows explicitly that reduced field scale factor $\alpha/2\pi$ for the electromagnetic field is almost cancelled by concurrent and purely classical recoil mass-field effect $2m_e/m_H \approx 2m_e/M_p$, an important effect for circular orbits only, *invisible* in Bohr or QED/QFT theory [12a]. This important connection is elaborated more in detail elsewhere [25] but it is obvious by now that this result is obtained from first principles only, just by introducing the self-energy of the H atom, determined by $m_H$, in the bound state theory as a datum to derive an extra equilibrium condition like (7n), overlooked by Bohr and in bound state Dirac-based QED.

(v) *A 4-unit charge structure for perturbed H: symmetry considerations and chirality*

Before using experimental data to verify theoretical results in (iii-iv) for circular Bohr orbits in H, we must discuss the symmetry of a 3D 4-unit charge Coulomb complex $F_{4H}$, which can be denoted as

$$F_{4H} : \{e_1, e_2; e_3, e_4\} \qquad (8a)$$

The shape of this structure is determined by the interaction between two linear 1D structures in (7a). A classical explanation for chiral behavior of $F_{4H}$ in line with the classical rules for chiral behavior in (i) requires that all 4 charges in (8a) be different. A simpler argument is provided by the fact that steps (7b) and (7c) produce a 1D line segment for each of the 2 terms in (7a). Then, the structure of complex (8a) corresponds with a spatial combination of 2 line segments $\lambda$ (7b) and r (7c), each with 2-unit charges at their vertices. As remarked above, this is of the required form to result, eventually, in a *3D structure* with 4-fermions or 4-unit charges at the 4 vertices. Instead of a classical chiral molecule ABCD in (ii), we get a generic 4-fermion or 4-unit charge structure (8a) for perturbed H [12a], elaborated analytically in [25]. Reminding that 4-particle systems are not soluble, the present generic analysis suffices to open the possibility that the 3D structure of intermediate complex $F_{4H}$ (3a) can be either[4]

---

[4] The physics behind the achiral and chiral character of linear 1D structures is given elsewhere [25].

a) of *achiral* type, when this 1, 2 or 3D 4-fermion structure $F_{4H}$ (8a) has at least *one symmetry element* or

b) of *chiral* type, when this single symmetry element has *disappeared* in structure $F_{4H}$. Only in the *chiral* case (b), different left- and right-handed forms $F_{4H(L)}$ and $F_{4H(R)}$ must show, preferentially detectable by means of a Mexican hat type PEC, as pointed out above.

Without elaborating here on the details of achiral or chiral forms of complex $F_{4H}$ [25], intra-atomic *parity-effects* will be needed to describe this left/right- or mirror symmetry as soon as it shows. If planar $F_{4H}$ behaves exactly as a classical *achiral* complex like the intermediary achiral 2D ABCD structure in a Walden inversion (5), we expect

(1) that the transition from $F_{4H(L)}$ to $F_{4H(R)}$ is characterized exactly by the same critical point (6), of which at least one generic property is *exactly* known, i.e.

$$F_{4H(planar\ or\ 2D)}\ (\text{critical point}) \sim \tfrac{1}{2}\pi \qquad (8b)$$

as argued around (6) in (i-ii) [12a,20] and

(2) that a double well or Mexican hat potential curve must be detected, probably a quartic, containing as highest power

$$PEC_{H \leftrightarrow \underline{H}} \sim 1/n^4 \qquad (8c)$$

We know from Noether's theorem [27] that *parity-effects* in dynamical fermion systems can be monitored by means of the *angular* momentum of fermions, subject to (circular) motion [28], an important tool we will use below. In Bohr's original $1/n^2$ theory, principal quantum number n is an integer. Knowing that, however, Bohr theory is not exact, we can attribute these errors to the fact that all n's are not exactly equal to an integer (see [12a] and further below).

*In this paragraph, we learned that classical chiral behavior, applied to perturbed H, leads to two very stringent quantitative criteria (8b) and (8c), which must be obeyed for any natural H↔$\underline{H}$ transition process like (1) or (4), if it exists.*

(vi) *Monitoring the behavior of the (a)chiral 4-fermion complex $F_{4H}$ in perturbed H for circular orbits ($ns_{\frac{1}{2}}$-states)*

In *achiral* Bohr $1/n^2$ theory, angular momentum, the critical Noether-variable [27] when it comes to assess parity-effects within $F_{4H}$, is described sufficiently with integer principal quantum number n. We call original Bohr $1/n^2$ theory *achiral*, as it is invariant to algebraic parity effects upon n, say to a switch from +n to –n: in Bohr $1/n^2$ *theory*, parity is always conserved. In this hypothesis, Bohr $1/n^2$ theory is an exact theory for species H in its intermediary (*virtual*) achiral state like $H_{achiral}$, introduced in (4). But in the limit, this achiral parity non-violating character seems to be the reason why Bohr $1/n^2$ theory is not exact: there are relatively small deviations with experiment even for the simplest case of circular orbits. In [12a], we already zoomed in on these relatively small errors. To test *achiral* Bohr $1/n^2$ theory with *integer n*-values for $ns_{1/2}$-states, we calculated Rydberg-values or *running* Rydbergs $R_H(n)$ for each state in the Lyman series with [12a]

$$-E_{nH}.n^2 = R_H(n) \qquad (9a)$$

These non-constant or *running* $R_H(n)$-values must be interpreted as *deviations from integer n* (see also [14,28] for an alternative discussion). With Noether, deviations of $R_H(n)$ from a constant value cannot but reflect intra-atomic parity-effects (*internal symmetries*) in perturbed H or in $F_{4H}$ (3a) [28]. Quantitatively, we obtained small (order $10^{-5}$) deviations from integer n (or better $n^2$) [12a] using quite accurate QED – $E_{nH}$ values, calculated a long time ago by Erickson [21]. Their accuracy is of order 0.000 001 $cm^{-1}$[21], close to the Hz limit for H-terms now in reach experimentally (see term H1S-2S in [11]). At least theoretically, this would give a *precision* for C in (3) of order parts in $10^3$.

There may be an uncertainty about the *phenomenological* interpretation of the results in [12a], since Erickson's data [21] are based upon bound state QED. Here, the asymptote connected with the deviations of Bohr theory finds its origin in the relativity correction, giving $\alpha^2 R_H$ $cm^{-1}$ (see above) or order $10^{-5}$ as required. In original QED, based upon the Dirac-Sommerfeld terms, critical $n_0$ for $ns_{1/2}$- as well for $np_{1/2}$-states is 1.5, a half-integer, leading to *a degeneracy of $ns_{1/2}$- and $np_{1/2}$- states* [23,28]. A critical point for bound state QED [23] is still the explanation to be given for the standard Lamb shift [29], clearly proving that *H $ns_{1/2}$- and $np_{1/2}$-states are not degenerate*. As shown above around (7n), a 4-unit charge model for *perturbed H* leads to a scaling effect for n by means of $\pi$ for *circular orbits* or H $ns_{1/2}$-states. The fact that Bohr's principal quantum number n may have to be scaled by irrational number $\pi$ also, see (7n) and [12a], is *a classical CSB result but also an absolute novelty in atom theory*. It is a possibility not considered in bound state QED, not even in its most recent forms [23]. To illustrate its impact, we go back in time, avoid QED-theory and use *experimental Lyman $ns_{1/2}$-terms* and confront these with criteria (8b) and (8c). In fact, we use a semi-empirical approximation for observed lines in the atomic spectrum, just like Bohr did when he referred to the regularities in the lines as observed a long time before him by Rydberg, Ritz….

(vii) *Experimental disclosure of classical critical point (3b) for chiral behavior of perturbed H or 4-fermion complex $F_{4H}$*

To avoid momentarily the burden of bound state QED and its sophisticated calculations [21,23], we proceed on a phenomenological basis and use the *best experimental terms $T_{nH}$ available for the Lyman $ns_{1/2}$-series*. These were tabulated by Kelly [22] and are shown in Table 1. Their precision is *only* 0.0001 cm$^{-1}$ (about 3 MHz). Using these terms, we can calculate the 20 level energies $E_{nH}$ using the Bohr formula

$$-E_{nH} = R_H - T_{nH} \qquad (9b)$$

reminding that Kelly's limit, the $R_H$-value, is 109678.7737 cm$^{-1}$ [22]. Running Rydbergs $R_H(n)$ in cm$^{-1}$ are obtained after multiplying (9b) with $n^2$ as in (9a)(see Table 1). Fig. 1 shows a plot of $R_H(n)$ versus $1/n$ [12a]. A simple quadratic fit

$$R_H(n) = = -4.42364/n^2 + 5.62565/n + 109677.570351 \text{ cm}^{-1} \quad (10a)$$

is fairly accurate (goodness of fit $R^2 = 0.9997$). Result (10a) may be less *accurate* than but is very similar to Eqn. (1) in [12a]. The Rydberg for $n=\infty$ is related to $R_\infty$ by the recoil correction $1/(1+m_e/M_p)$, giving $R_\infty = 109737.3026$ cm$^{-1}$. Maximum $R_H(n)$ can be called the *harmonic* Rydberg [12a] and is equal to

$$R_{harm} = 109679.35892 \text{ cm}^{-1} \qquad (10b)$$

on the basis of Kelly-data and fit (10a). The QED-data based value for $R_{harm}$, derived in [12a], is 109679.352282 cm$^{-1}$, a difference of order 10 MHz with (10b). Part of the numerical differences with the QED-based results in [12a] and those reported here is due to the difference in the constants used by Erickson and Kelly to get at energies in cm$^{-1}$.

Interesting applications of a harmonic Rydberg $R_{harm}$ are given elsewhere [28] and will be illustrated further below. As in [12a], we remark that this critical Rydberg is not given by NIST [26] and has never played a significant role in bound state QED either [21,23]. Nevertheless, *it must have an important classical meaning too* [12a]. In fact, we can now associate this harmonic Rydberg tentatively with the intermediary state for the *achiral* state of species H, appearing in (4) but missing in (1).

In view of the internal symmetries hidden in processes (1) and (4), the outcome for the *internal* structure of species H based upon *experimental* data [22] for its Lyman series is the more important. If this symmetry is really internal as we suspect, this result based upon less accurate experimental data [22] must, quantitatively, be very close to that obtained with *QED-one-electron energies* [21], *despite the difference in accuracy*. With Kelly data [22] and (10a), an extremum for the parabola in Fig. 1 is obtained at

$$n_0 = 1.5727 \approx \frac{1}{2}\pi \qquad (11)$$

close to the theoretical critical n-value (6) or (8b) we anticipated for a classical 19th century Walden-type inversion between left- and right-handed 4-fermion structures for species H in (4) as discussed above. In [12a], critical $n_0$ is 1.5723, a difference of only $2.5 \cdot 10^{-4}$ with (11) based upon less accurate *experimental* data [22]. *The use of less accurate data [5] does not prevent a consistent and successful disclosure of the very same internal mirror symmetry in H as already detected in* [12a]. This surprisingly consistent result makes it difficult to understand why there is an explicit warning [30] that Kelly-data *have not (yet) been evaluated by NIST*.

(viii) *The consistency of the predictions (6), (8b) and (7n) for critical complex $F_{4H}$ with Bohr theory*

Clearly, the precision of spectral data (terms) is not that decisive to disclose internal symmetries in species H. Less accurate terms [22] lead to the same end result (11) as more accurate QED-data [21]. Both lead to the same critical point for chiral behavior as in a classical 19th century Walden inversion (5) and (8d). Nevertheless, how important result (11) may be in the context of intra-atomic *mirror symmetry* and *chiral behavior* for H, experimental result $n_0=\frac{1}{2}\pi$ in (11) is wrong by a factor of 2 when compared with theoretically expected result (7n) or $n=\pi$. The latter is based upon our hypothesis of a 4-fermion structure for perturbed H and *the energy implications* of (7a), among which the effect of incorporating the hydrogen self-energy $m_H c^2$ in our model.

We are obliged to find the effect of the mirror symmetry of intermediate chiral structure $F_{4H}$ expressed by (11) on the level energies. As outlined in [12a] and using (11) analytically, the parabola in Fig. 1, obeying (10a), can be rewritten more generally as

$$R_H(n) = A + B(1-\tfrac{1}{2}\pi/n)^2 \qquad (12a)$$

Using Bohr theory, the effect of *intra-atomic mirror symmetry* (11) in chiral H on its *level energies* must be

$$-E_{nH} = R_H(n)/n^2 = A/n^2 + B(1-\tfrac{1}{2}\pi/n)^2/n^2 \qquad (12b)$$

---

[5] Even using less accurate observed terms (rounded at 0.001 cm$^{-1}$, error of 30 MHz), critical n is 1.568, still close to ½π but *lower*. Since (11) is *larger* than ½π, number ½π might be extremely important role for high precision terms.

whereby the Kelly value for A (=109678.7737 cm$^{-1}$) and B=0 produce a Bohr result[6] of type $A/n^2$. This potential (12b) is fourth order $1/n^4$, as mentioned around (8c). It will be discussed further below in connection with Mexican hat type potentials, in line with the observed terms of the H-spectrum. The extremum for level energies *for achiral circular orbits* can be obtained analytically from (12b), by requiring that $-dE_{nH}/dn=0$, which gives

$$n = \pi \qquad (13)$$

*This analytical result (13) is exactly the hidden or invisible result (8n) derived above (7n) in the hypothesis of a chiral 4-fermion structure $F_{4H}$ for perturbed H.* The only logical interpretation for *chiral behavior of* intermediate 4-fermion complex $F_{4H}$ in *perturbed H* is

$$F_{4H(Left)} \leftrightarrow F_{4H}\,(\tfrac{1}{2}\pi) \leftrightarrow F_{4H(Right)}$$

This means that the natural H↔H transition is indeed of classical 19th century Walden-type (5)

$$H \text{ [or } H_{Left}\,(H_{Right})] \leftrightarrow H_0(\tfrac{1}{2}\pi) \leftrightarrow \underline{H} \text{ [or } H_{Right}\,(H_{Left})] \qquad (14)$$

more detailed than (4) and a considerable improvement over standard conversion process (1). As suggested in [12a], this directly leads to a natural and observable hydrogen-antihydrogen conversion and to an alternative but very *classical* mechanism for transition (1).

Generic and classical results (6), (7n), (8b) and (11) were all anticipated on the basis of the equally generic character of a purely classical 19th century left-right or mirror *atom-antiatom symmetry* [12]. These quantitative results are generated formally by respecting the importance of and by introducing the hydrogen self-energy into the theory for hydrogen, which now seems to be only very normal and natural, if not even trivial, a procedure. *It is difficult to imagine why this very simple and elegant procedure, based upon (7a), was not followed earlier.*

*(ix) Conclusive evidence: a Mexican hat curve in natural perturbed species H*

We are now left with providing conclusive evidence for our thesis that natural perturbed H shows chiral behavior indeed and that the H↔H conversion process (4) occurs naturally. To do so, we subtract Bohr-type *achiral* $R_{harm}/n^2$-values from the *observed* level energies (or better terms), given by (9). These differences

---

[6] For terms, the errors of Bohr theory (parts in $10^7$) are decreased (parts in $10^9$) using semi-empirical result (12).

$$\Delta_{nH} = R_{harm}/n^2 + E_{nH} \qquad (15)$$

represent the phenomenological so-called *chiral effects in the perturbed H-system* as they can be read from the *observed spectrum of species H* available for many a decade. The values are also collected in Table 1 (last column). In Fig. 2 these values are plotted versus variable (π/n-1), which represents, completely in line with our derivations above, the deviations of principal quantum number n from scale factor π for circular orbits. As we cannot but expect this to be a genuine field effect (see Introduction), the result is a perfect Mexican hat shaped PEC, fully in line with the Hund analysis of classical chiral behavior. We notice, however, that the fit only produces the complete Mexican hat curve, if the curve for the data is (artificially) extrapolated to the left, i.e. to n-values difficult to understand (but see [20d] for a classical explanation).

Nevertheless, we notice that the maximum in between the wells for the left- and right-handed H-forms is very close to π (since π/n-1=0 or n= π), as expected theoretically (7n) and (13). This confirms the validity of our starting point, equation (7a) with H self-energy included, to describe the perturbed species H. We stress that this PEC is obtained with *observed* (Kelly) term values, not with Erickson's QED level energies, which we used elsewhere to extract a very similar double well potential [28].

The remarkable thing is that observed term-values are known since many a decade. However, to the best of my knowledge, this classical chiral analysis of *the observed spectrum of species H*, leading to Fig. 1 and 2, has never been reported by others.

Having overlooked this essential and available information on the reality of H and H̲-forms in nature, as well as the complete inability to get a better theoretical understanding of the chiral behavior of species H, has eventually led to the very complex experiments, now being performed [6-9], and under review here. Of course, Fig. 1 and 2, being based upon a CSB-theory as well as on observed data, must have an impact on the status of bound state QED, as we stated frequently [12a].

At a different level of aggregation (say chemical bonds), the possibility that atom-antiatom switches are natural and can be observed in nature, must also have repercussions for the status of the theory of the chemical bond, as we remarked earlier at a number of instances (see [12b] and references therein).

We admit that Bohr was probably unable to deal with these essential details of the H-spectrum due to the limited precision of the spectral data available at the time. At least a precision of 0.001 cm$^{-1}$ is required[7] for the Lyman terms to get at the details shown in Fig. 1 and 2.

(x) *Some theoretical consequences*

In [12a] we argued that the framework of bound state QED cannot be validated as it stands today, since it has no simple and straightforward explanation for the fact that irrational number $\pi$ may play a critical role for scaling principal quantum number n in a chiral theory for atomic spectra. In bound state QED for $p_{1/2}$- and $p_{3/2}$-sates, the critical n-values are rational (half-integer or integer) numbers 1.5 and 3 respectively [28]. Standard Lamb shifts [29] disprove the theoretical degeneracy of $s_{1/2}$- and $p_{1/2}$-states originally expected from old QED. This degeneracy was removed mainly on account of new terms introduced in the theory (such as Bethe logarithms, which are much larger for $ns_{1/2}$- than for $np_{1/2}$-states [23]). Still, even in the theoretical framework of modern bound state QED [21,23], there is no direct theoretical relation between number $\pi$ and quantum number n for $ns_{1/2}$-states either. From our phenomenological interpretation of observed Lyman terms (see Fig. 1 and 2 and also [12,28]), the natural appearance of $\pi$ must be understood in terms of scaling for (circular) orbits, either larger or smaller than a given unit [14,28,30]. These interesting but unforeseen theoretical implications of the present analysis and those in [12a] for equations (1)-(3) as well as for atomic physics in general are dealt with elsewhere [14,30]. It is obvious that there must be an additional criterion for equilibrium in atomic systems, overlooked in Bohr theory as well as in QED. This extra condition leads to result (7n) directly from first principles as collected in the simple formula (7a). We now see that the persistent neglect of hydrogen self-energy $m_H$ as a major anchor for the H theory is responsible for failing to recognize that there may also be a linear connection between n and $\pi$ for 1D, at variance with that put forward by de Broglie. Whether or not this will be sufficient an explanation to understand Lamb-shifts will have to be decided at a later instance (the near degeneracy of field scale factors $2\pi/\alpha$ and $2m_e/M_p$, mentioned above, may be another element in this evaluation).

---

[7] The reader may easily verify the generation of the different parabola in Fig. 1 by rounding the Kelly –EnH values in Table 1 gradually from 0 to 0.0001.

We hope our analysis may result in a significant simplification of bound state QED (see also [30]), if not in *a closed form bound state theory*, we all badly want. The use of static linear structures (line segments) instead of dynamic circular structures to analyze the bound electron-proton system as suggested in this report may lead to a better understanding of this fundamental interaction and the internal symmetries hidden in conversion process (1) and (4) [30]. As stated above, an essential problem lies in the interpretation of angular and radial velocities in central force systems like in neutral atoms [14] and in the meaning of the necessity to incorporate the self-energy of neutral species H.

With respect to the theory of chemical bonding, the results of an atom-antiatom switch in the molecular Hamiltonian leads to a great simplification also, theoretically as well as computationally [12b]. The observed shape invariance of most molecular PECs and their singlet-triplet splittings is fully in line with this parity operator within the molecular Hamiltonian [12b].

*(xi) Conclusion*

*The mere fact that not only classical generic result (7n) is detected in the observed line spectrum of natural H but also that a Mexican hat curve is hidden therein, strongly supports our point that the mirror-version of atom H, an antihydrogen or antiatom H, is already provided by nature itself.* We proved that the H↔H transition (1) really appears in *perturbed* H and that it can be described with the same procedure as applying to a classical 19th century Walden inversion and in line with Hund's early analysis. In terms of a trigonal pyramid model for *perturbed H* (7a) we would get a Walden inversion of *fermionic* type

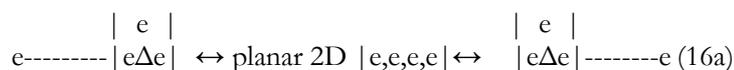
$$\text{e}\text{---------}|\text{e}\Delta\text{e}| \leftrightarrow \text{planar 2D } |\text{e,e,e,e}| \leftrightarrow |\text{e}\Delta\text{e}|\text{--------}\text{e} \quad (16a)$$

which is a more detailed description for a natural H-H transition (1)

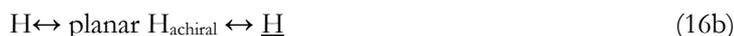
$$H \leftrightarrow \text{planar } H_{achiral} \leftrightarrow \underline{H} \qquad (16b)$$

In nature, there is very subtle procedure indeed to impose left-right characteristics on neutral particle systems, *whereby it is essential that the neutral particle's self-energy must be taken into account* (see also [14b]). Results (14) and (16) explicitly support our thesis above that recent claims [6-9] are probably premature, since an intra-atomic mirror symmetry effect may already be hidden in natural perturbed H: the observed fine structure in atomic line spectra derives from a hidden classically understandable *intra-* *atomic mirror asymmetry*. This purely classical description was overlooked, left unnoticed or interpreted differently in the framework of QFT (QED), as argued in [12a,25]. Even the less accurate *experimental* Kelly data [22] confirm these earlier findings [12a] and we repeat that this generic result is consistent with the seminal work of Lamb and Retherford [29] more than half a century ago. The origin of the standard Lamb shift may reduce finally to the small difference between number 1.5 for non-circular and (irrational) ½π for circular orbits.

If result (16) is valid for natural H, the mirror symmetrical H̲ version is already contained *in nature*. This contradicts the *premature* claims [6-9] that mass-production of *artificial* H̲, i.e. structure $e^+,p^-$, is needed to assess the critical interval H̲1S-2S [12a,25] and to test the CPT-theorem. The corresponding parity-violating energy shifts for H (*not yet given* in Table 1 and *invisible* in Fig. 1 and 2) can be derived *quantitatively* from the available line spectrum of natural H. These results are presented elsewhere [28,31], where we will also compare formally and in detail the classical approach used here with the framework of bound state QED. We believe it is not premature to say that (significant parts of) both the H- and the H̲-line spectra are assessable from that of natural *species* H [31]. There is only one line spectrum in nature for a neutral 2-unit charge structure obeying Z=1. This is a naturally occurring neutral species called hydrogen or H and has exactly the chiral properties described in this work. All this may have pertinent consequences for the general theory for (1), for natural symmetry breaking, for chiral behavior, for baryon- and lepton-number conservation, for matter-antimatter asymmetry [30], for a low energy Higgs mechanism,… as well as for ongoing experiments like [6-9]. The existence of natural H̲-states confirms our earlier deductions made on the basis of molecular spectra [12b]. Further quantitative criteria are presented elsewhere.

We thank G.W. Erickson, C. Chantler and B. Crasemann for contacts on the data problem and D. Avnir, J.D. Dunitz, and L. Wolniewicz for issues related to chirality. We are in debt to B. Sutcliffe for instructive talks.

Table 1. One-electron energies $E_{nH}$, $R_H(n)$-values from Kelly term values $T_{nH}$ (in cm$^{-1}$) [22] for the H

Lyman ns½-series

| n | 1/n | $T_{nH}$ | $-E_{nH}$ | $R_H(n)$ | $R_{harm}/n^2+E_{nH}$ |
|---|---|---|---|---|---|
| 1 | 1.0000 | 0.0000 | 109678.7737 | 109678.7737 | 0.585220 |
| 2 | 0.5000 | 82258.9559 | 27419.8178 | 109679.2712 | 0.021930 |
| 3 | 0.3333 | 97492.2235 | 12186.5502 | 109678.9518 | 0.045236 |
| 4 | 0.2500 | 102823.8549 | 6854.9188 | 109678.7008 | 0.041133 |
| 5 | 0.2000 | 105291.6329 | 4387.1408 | 109678.5200 | 0.033557 |
| 6 | 0.1667 | 106632.1518 | 3046.6219 | 109678.3884 | 0.026959 |
| 7 | 0.1429 | 107440.4413 | 2238.3324 | 109678.2876 | 0.021864 |
| 8 | 0.1250 | 107965.0517 | 1713.7220 | 109678.2080 | 0.017983 |
| 9 | 0.1111 | 108324.7225 | 1354.0512 | 109678.1472 | 0.014960 |
| 10 | 0.1000 | 108581.9928 | 1096.7809 | 109678.0900 | 0.012689 |
| 11 | 0.0909 | 108772.3435 | 906.4302 | 109678.0542 | 0.010783 |
| 12 | 0.0833 | 108917.1208 | 761.6529 | 109678.0176 | 0.009315 |
| 13 | 0.0769 | 109029.7916 | 648.9821 | 109677.9749 | 0.008189 |
| 14 | 0.0714 | 109119.1923 | 559.5814 | 109677.9544 | 0.007166 |
| 15 | 0.0667 | 109191.3163 | 487.4574 | 109677.9150 | 0.006417 |
| 16 | 0.0625 | 109250.3444 | 428.4293 | 109677.9008 | 0.005696 |
| 17 | 0.0588 | 109299.2655 | 379.5082 | 109677.8698 | 0.005153 |
| 18 | 0.0556 | 109340.2618 | 338.5119 | 109677.8556 | 0.004640 |
| 19 | 0.0526 | 109374.9569 | 303.8168 | 109677.8648 | 0.004139 |
| 20 | 0.0500 | 109404.5791 | 274.1946 | 109677.8400 | 0.003797 |

Fig. 1 Plot of running Rydbergs $R_H(n)$ versus $1/n$ for the H Lyman-series

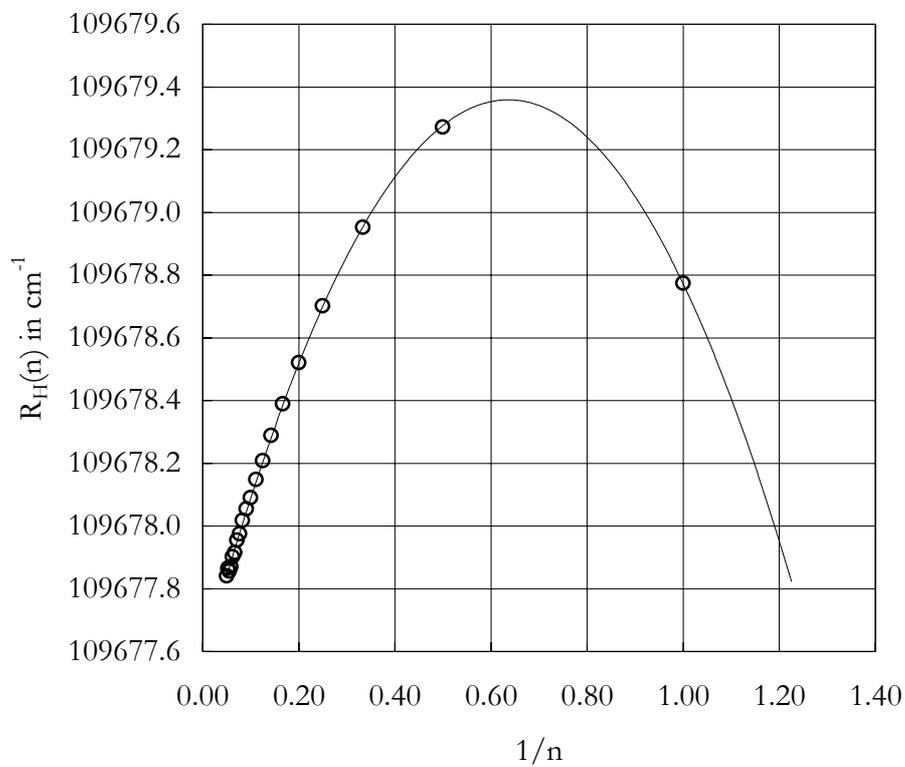

Fig. 2 Mexican hat curve extracted from observed Kelly data for the Lyman ns-series

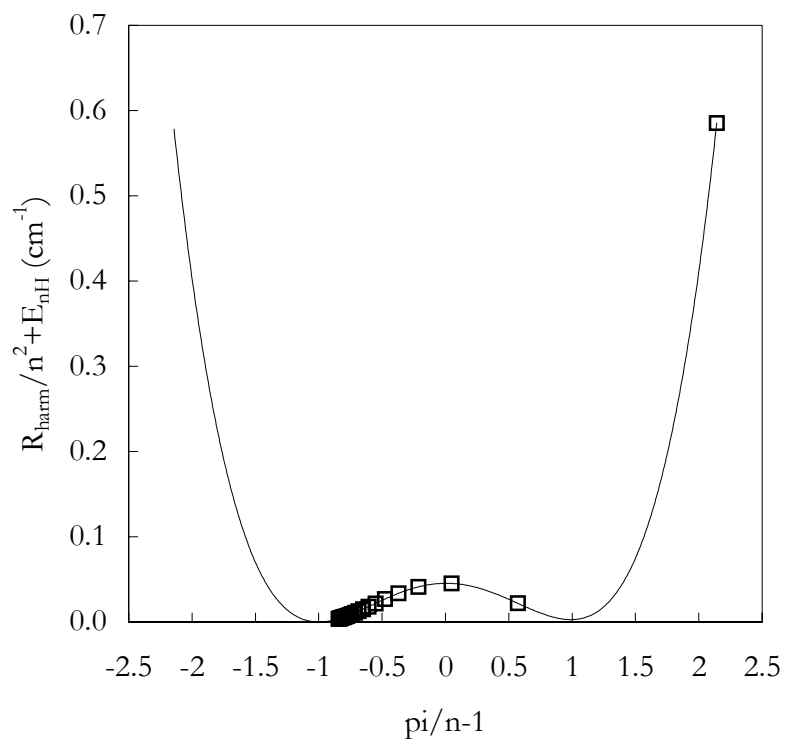

curve fit y= $0.04482x^4 + 0.00018x^3 - 0.08879x^2 + 0.00119x + 0.04539$ cm$^{-1}$